\documentclass{sig-alternate-05-2015}
\usepackage{epstopdf}
\usepackage{multirow}
\usepackage[table]{xcolor}
\usepackage{array}
\usepackage{dblfloatfix}
\usepackage{subfig}
\usepackage{url}

\begin{document}
%
\conferenceinfo{WOODSTOCK}{'97 El Paso, Texas USA}
\CopyrightYear{20XX} 



\title{Trustworthiness in Enterprise Crowdsourcing: a Taxonomy \\ \& evidence from data\thanks{Author's submitted version.
 Accepted at ICSE SEIP 2016. Published version at: https://dl.acm.org/citation.cfm?id=2889225. DOI: http://dx.doi.org/10.1145/2889160.2889225. Copyright: ACM}}

%
%
%
%
%

\numberofauthors{4} 
%
\author{
%
%
\alignauthor
Anurag Dwarakanath\\
       \affaddr{Accenture Technology Labs}\\
       \affaddr{Bangalore, India}\\
       \email{anurag.dwarakanath\\@accenture.com}
\alignauthor
Shrikanth N.C.\\
       \affaddr{Accenture Technology Labs}\\
       \affaddr{Bangalore, India}\\
       \email{shrikanth.n.c\\@accenture.com}
\alignauthor Kumar Abhinav\\
       \affaddr{IIIT-Delhi}\\
       \affaddr{Delhi, India}\\
       \email{kumar1365@iiitd.ac.in}
\and  
\alignauthor Alex Kass\\
       \affaddr{Accenture Technology Labs}\\
       \affaddr{San Jose, USA}\\
       \email{alex.kass@accenture.com}
}

\maketitle
\begin{abstract}
In this paper we study the trustworthiness of the crowd for crowdsourced software development. Through the study of literature from various domains, we present the risks that impact the trustworthiness in an enterprise context. We survey known techniques to mitigate these risks. We also analyze key metrics from multiple years of empirical data of actual crowdsourced software development tasks from two leading vendors. We present the metrics around untrustworthy behavior and the performance of certain mitigation techniques. Our study and results can serve as guidelines for crowdsourced enterprise software development. 
\end{abstract}




\section{Introduction}
Crowdsourcing is an emerging trend where a group of geographically distributed individuals contribute willingly, sometimes for free, towards a common goal. Many successful examples of crowdsourcing have been witnessed - from the large scale annotation of data \cite{alonso2009can} to the design of an amphibious combat vehicle \cite{tank}.
%
%

In this paper, we study the application of crowdsourcing in the context of enterprise software development. Crowdsourcing has a number of potential benefits over the existing software development methodologies \cite{Stol2014}\cite{dwarakanath2015} including: a) Faster time-to-market due to parallel execution of tasks; b) Lower cost due to access to the right skills; c) Higher quality through the creativity \& competitive nature of the crowd. 

A key issue with crowdsourcing is the apparent lack of `trust' on the crowd - i.e. what is the guarantee that the crowd would not jeopardize a task? There are numerous examples where crowdsourcing has been troublesome. A prominent example is the DARPA shredder challenge \cite{stefanovitch2014error}. This challenge required putting together shredded documents. Teams that participated in the challenge, included those pursuing algorithmic approaches as well as a crowdsourcing based approach. The crowdsourcing based team began well, beating some of the algorithmic approaches. However, one of the crowd workers turned `rogue' and sabotaged the crowdsourcing initiative by repeatedly undoing the work done by other crowd workers. Eventually, the motivation of the legitimate crowd workers was eroded and the team failed to complete the challenge. The saboteur later revealed \cite{loneHacker} that he was part of a competing algorithmic based team and wanted to show some of the severe consequences of crowdsourcing. 
\begin{figure} [b!] 
\centering
\epsfig{file=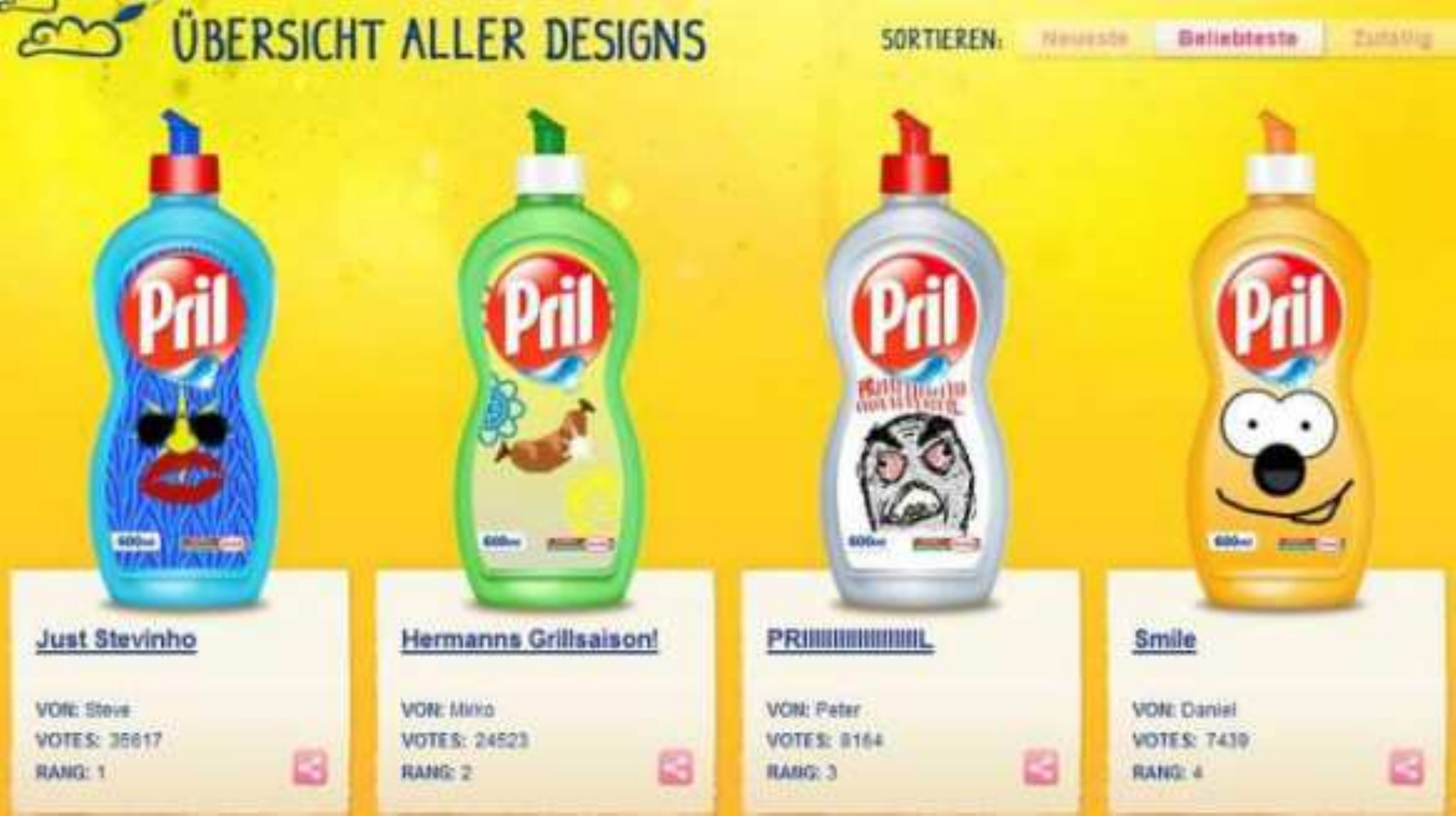, width=2.9in}
\caption{The most voted designs from the crowd.}
\end{figure} \label{prilFig}

A similar sabotage was seen in the social media campaign of Henkel where the crowd was asked to design and rate a packaging image for one of its products \cite{pril}. As a prank, some of the crowd workers submitted humorous designs. These designs also got voted to the top ranks (see Figure \ref{prilFig} for the most voted designs). The company had to retract the public voting scheme and put in place a set of internal reviewers to choose the best design. The crowdsourcing initiative resulted in bad publicity defeating the very purpose it was set out to achieve.

Crowdsourcing of software development tasks have also shown numerous problems. We looked at the comments from the crowdsourcers for tasks posted on a leading vendor (Upwork\cite{upwork}). One such comment is shown below. More comments are available in Table \ref{comments}. 

\begin{quote}
	\itshape{``This guy is a scumbag. Did[n't] do any work and wasted my time. Then begged for money when I ended the contract. When I refused a virus ended up on my site.''}
\end{quote}

In this paper, we study the risks involved in enterprise crowdsourcing.  Our work is situated in the context of an enterprise getting its software development tasks completed through the crowd. However, our work is also applicable in the context of crowdsourcing creative \& subjective tasks (like the design of a logo). Most existing studies on the trust of the crowd have been limited to the context of data annotation and look at only the quality of the deliverable from the crowd. Through a study of literature from varying domains (including legal, security, multi-agent systems) and our experience through crowdsourcing experiments, we present a taxonomy of trustworthiness which goes far beyond the basic attribute of quality. We also survey known risk mitigation techniques. Finally, we analyze a large set of data from actual software development crowdsourcing tasks from two leading vendors - Upwork (formerly oDesk) \cite{upwork} and  TopCoder \cite{topcoder}. We present results of this analysis and the performance of certain mitigation techniques. 

We believe, our work will advance the study of crowdsourcing through a few key contributions - a) a comprehensive taxonomy of trustworthiness; b) existing techniques to mitigate the risks; and c) examples of real-life issues through crowdsourcer's comments; d) statistical evidence of the performance of certain mitigation techniques.

This paper is structured as follows. We present the related work in Section \ref{relatedwork}. The taxonomy of trustworthiness is presented in Section \ref{taxonomy}. We present the various existing techniques in Section \ref{existingTechniques}. The empirical evidence from data is in Section \ref{evidence} and we conclude in Section \ref{conclusion}.

\section{Related Work} \label{relatedwork}

As a preliminary, we will introduce the basic mechanism of crowdsourcing and the terminology that will be used in this paper. A crowdsourcing effort involves three stakeholders - a) \textit{the crowdsourcer}: This is the entity which posts a task to be completed. The crowdsourcer specifies the task description and supplies any required data \& tools; b) \textit{the crowd}: This is the entity which performs the task. Individuals in the crowd are denoted as a crowd worker and the deliverable from the crowd worker is denoted as a work product; c) \textit{the crowdsourcing platform}: This is the platform (or vendor) which brings together the crowd and the crowdsourcer. There are two broad types of crowdsourcing tasks: Micro-tasks which require a small amount of effort to complete and does not require specialized skills; Macro-tasks which requires a lot more effort and need specific skills. We now explore the related work in trustworthiness.

The notion of trust has been traditionally studied in the context of multi-agent systems such as e-commerce platforms (how do we select the right vendor for a product), online communities (how do you know a review of a product is not biased) and semantic web (how do you choose the right service provider). In the context of crowdsourcing, we adapt the definition of trust from \cite{Huynh2006}.
\begin{quote}
	Trust is the belief that a stakeholder will perform his duties as per the expectation.
\end{quote}



Trustworthiness research thus looks to develop mechanisms such that the `belief' in the performance of a stakeholder can be built. In the context of enterprise crowdsourcing, we are focused on building mechanisms to establish the belief in the actions of a crowd worker. 


Work that studies the factors impacting trustworthiness has been limited to studying a single factor - the quality of work submission from the crowd. The work in \cite{allahbakhsh2013quality} proposes that the quality of the work product is influenced by the profile of the crowd worker and the design of the task. Their work also presents existing mechanisms to judge the overall quality of a submission.

The work in \cite{Eickhoff2011} studied the quality of annotation tasks in Amazon Mechanical Turk (AMT). They found that 38\% of crowd workers provide untrustworthy (i.e. poor quality) responses. Upon accepting only responses from the  `U.S.' geography, they found that the amount of untrustworthy responses fell by 71\%. This work concluded that choosing crowd workers based on origin is a strong way to improve trustworthiness.

A similar study on untrustworthy workers in open ended tasks was made in \cite{Gadiraju2015}. Their work showed that 43.2\% of the workers are untrustworthy and the geography of worker origin makes a big difference. This work also provided insights into the behavior of such workers. Most untrustworthy workers were found to be driven by monetary rewards and attempted to maximize their payoff by quickly completing as many tasks as possible; often at the expense of quality. Similar monetary minded motivation of crowd workers was seen in \cite{kittur2008crowdsourcing} and \cite{sorokin2008utility}.

The work in \cite{Ye} presented a model which recommends crowd workers for tasks based on their reputation. The work proposes that `satisfactory' results from crowd workers can be obtained by recommending those crowd workers who have completed tasks of similar type and similar rewards. The efficacy of the method was measured through simulation. 

Most of the related work focuses on only the quality of the submission. In our work, we present a comprehensive taxonomy of trustworthiness going beyond the basic attribute of quality. Further, most related work looks at micro-tasks based crowdsourcing, while our work specifically focuses on macro-tasks. We also provide empirical evidence from a large set of data from actual crowdsourcing experiments. 

\section{A Taxonomy of Trustworthiness in Crowdsourcing} \label{taxonomy}

In this section, we present the factors that influence the trustworthiness of the crowd. Our context is the development of enterprise software where the enterprise takes the role of the crowdsourcer and posts software development tasks. While the taxonomy is focused on software development, it can be generalized to macro-tasks as well.



\begin{figure} [b!] 
\centering
\epsfig{file=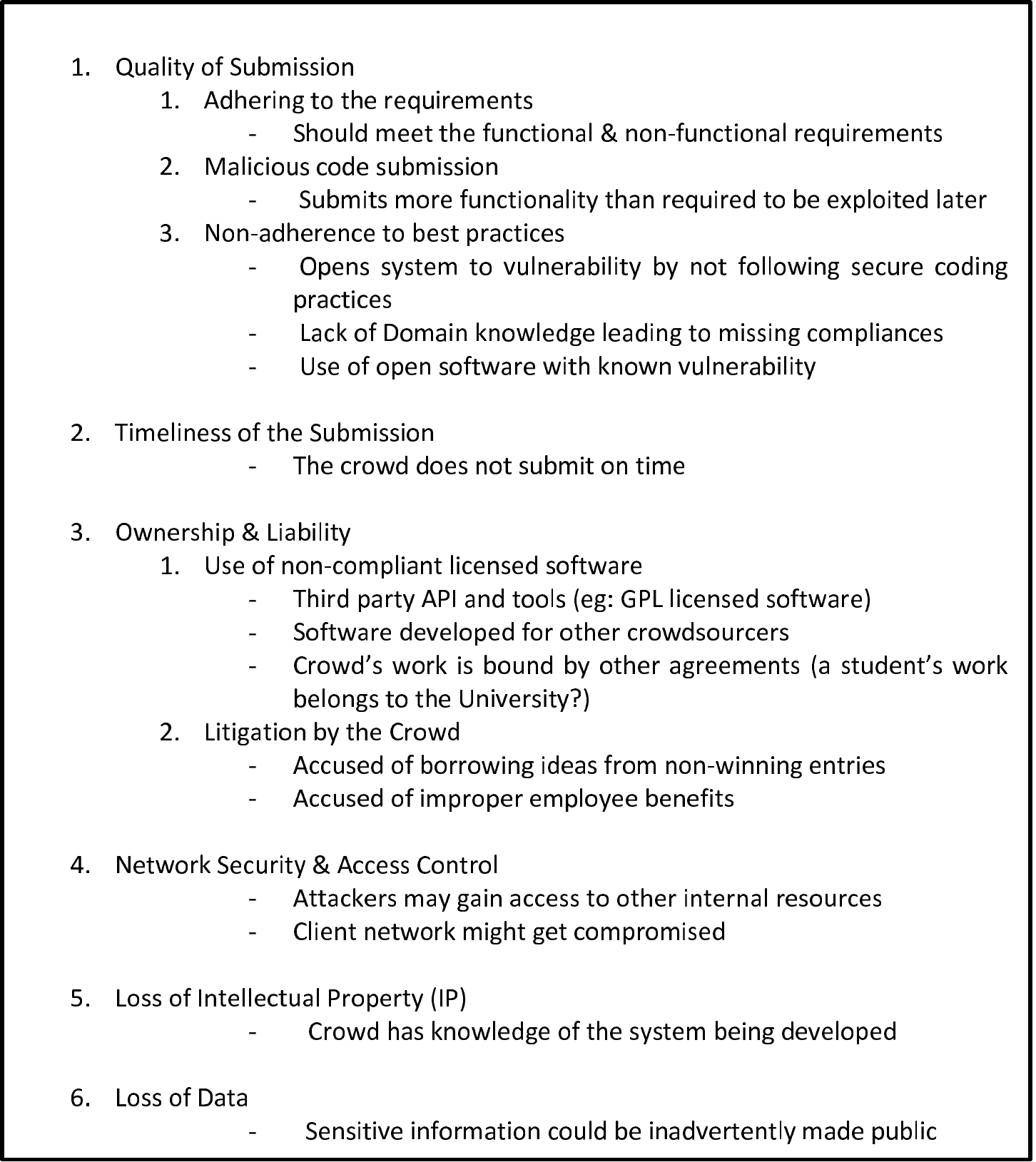, width=3in}
\caption{Taxonomy of Trustworthiness in Crowdsourcing.}
\label{tax}
\end{figure}

The taxonomy of trustworthiness in crowdsourcing is shown in Figure \ref{tax}. We explain each aspect below.

\subsection{Quality of the Submission}
A primary consideration with crowdsourcing is the quality of the work product. In the context of the implementation of software, the code that is submitted by the crowd can take the following quality attributes.

\subsubsection{Adhering to the requirements}
The crowd's work product is expected to meet the task requirements as specified by the crowdsourcer. The instructions typically include the functional \& non-functional requirements of the software. Not meeting the stated requirements is a conscious act of the crowd worker. It has been reported \cite{Gadiraju2015} that crowd workers look to complete a task quickly and try to minimize the amount of effort being spent on a task. This motivation could lead to cutting corners where certain requirements are not met (e.g. inefficient code which does not meet the non-functional requirements; poor exception handling; not considering boundary cases, etc.). 

\subsubsection{Malicious code submission}
Submission of malicious code is an extreme concern in crowdsourcing \cite{probCrowd}. We define `malicious' as those implementations that are performing more than the required functionality. The malicious code inserted could be exploited later when the software is deployed. Insertion of malicious code is a conscious act of the crowd worker.

\subsubsection{Non-Adherence to best practices}

This category captures those cases where the implementation may not follow the coding practices specified by the crowdsourcer or those typically followed in the industry. Examples of such practices include the naming conventions of methods and variables, appropriate comments and having an appropriate set of unit test cases. Missing to follow such best practices (when explicitly mentioned in the task description) mounts to a conscious action of the crowd worker. The underlying motivation for such an action could be driven due to the dynamics of the crowdsourcing environment where the amount of time given is small, rewards are strongly linked with the number of tasks completed and code review mechanisms are perceived to be weak. 

Non-adherence to best practices can also occur when the crowd worker is new to a particular domain and is unaware of the typical practice in the domain. For example, data regarding personally identifiable information is expected to be masked in an application that follows the HIPAA standard. Not being aware of such requirements could lead a crowd worker to unintentionally miss following the best practices. 

Finally, usage of software with known vulnerabilities is another factor to be considered. Here the crowd worker may leverage readily available software to complete a task. However, the software may have known issues which can be exploited by others. Such actions from the crowd worker may not originate due to a malicious intent.

\subsection{Timeliness of the Submission}
Crowdsourcing is characterized by the crowd's voluntary participation \cite{dwarakanath2015}. Thus, there is a possibility that no crowd worker volunteers to complete a task. The reasons could include poor monetary benefits \cite{sorokin2008utility}, poorly documented tasks \cite{probCrowd} or tasks requiring a large amount of effort. The formulation of the crowdsourcing tasks is the prerogative of the crowdsourcer and it is important to create tasks of the right granularity and with the right effort-benefit ratio \cite{probCrowd}. 

Another aspect of timeliness is where the crowd worker accepts a particular task and fails on his commitment. Such a scenario is particularly applicable in the context of Upwork where freelancers are hired. 

\subsection{Ownership \& Liability of the Submission}
The ownership of the crowd's submissions are expected to be transferred to the crowdsourcer, making its usage in enterprise applications amenable. However, there are a few considerations that have to be specifically verified from the point of view of the enterprise, as we elaborate below. 

\subsubsection{Use of non-compliant licensed software}
This category captures those cases where the crowd might use an open-source implementation that is bound by certain licenses not compliant for use in an enterprise. For example, GPL licenses are not amenable for commercial usage. 

A similar case is that of re-using one's own code which was developed for a different crowdsourcer and thus the rights of the code do not rest with the crowd worker. Such usage raises liability concerns due to conflict in ownership claims among the two crowdsourcers. 

Issue of non-compliant software also occurs when the crowd worker has signed agreements that give ownership of his creative work to other parties. An example of this case is that of a student studying in a university \cite{probCrowd}. Typically, while joining the university, the student would have agreed to a set of clauses that binds all creative work (like the implementation of software) to the university. The important aspect is that this handing-over of the rights to the university may not be even known to the student. Thus, as the student accepts to perform a crowdsourcing task, he may unintentionally create a liability for the crowdsourcer. 

The underlying motivation in the first two cases is typically explicit, where the crowd worker is aware of the fallacies, however the characteristics of the crowdsourcing engagement model (time, rewards \& review) may lead to such a behavior. The last case is typically unintentional.

The enterprise's use of the crowd's work can increase its exposure to infringement liability and other damages \cite{liberstein2012crowdsourcing}. Further, since the crowd worker typically lacks the assets \& resources to indemnify an enterprise \cite{probCrowd}, it would be the prerogative of the enterprise to ensure the submission would not attract legal claims and damages. 
 
\subsubsection{Litigation by the crowd}
The law governing crowdsourcing is still vague. Potential conflicts can arise around the claim of ownership of the creative works of the crowd worker and the role of the crowdsourcer as an employer. As an example of the former case, consider a scenario where two submissions (A \& B) are made to a task supplied by a crowdsourcer. Submission A is chosen by the crowdsourcer and the crowd worker is appropriately rewarded. However, at a later time the crowd worker who submitted work B can accuse the crowdsourcer of using his idea in the application. Since the crowdsourcer has viewed the work of B, this claim is legitimate. Thus, non-winning entries may increase the liability of the crowdsourcer \cite{probCrowd}.

Crowdsourcing also opens up an ambiguity as to who qualifies as an `employee' \cite{liberstein2012crowdsourcing} \cite{wolfson2011look}. The laws vary among the different geographies, but typically are based on the amount of control an employer has over the worker. The crowd worker may use a local law to accuse the crowdsourcer of improper employee rights \cite{wolfson2011look} (like the lack of a social security or the lack of health benefits). 

\subsection{Network Security \& Access Control}
To facilitate application development (e.g. to check-in code and run unit tests), the crowdsourcer typically provides the necessary access to the enterprise network. This access may allow a crowd worker to legitimately get inside the firewall and subsequently access sensitive information. Further, it has been suggested \cite{dwarakanath2015} that access control should be used to provide specific need based access. 

\subsection{Loss of Intellectual Property (IP)}
When the crowdsourcing approach is adopted, the crowdsourcer has willingly made public the requirements and the intended functionalities of the software product being developed \cite{probCrowd}. In particular cases where the intended software application is expected to bring competitive advantage, willingly giving away such an advantage even before implementation could lead to a serious business consequence. 

\subsection{Loss of Data}
When a crowdsourcer shares data with the crowd, there is a possibility of loss of sensitive information unknowingly \cite{wolfson2011look}. For example, AOL released about 20 million search queries from around 650000 users to the crowd for research purposes. Care was taken to anonymize the usernames and IP addresses. However, it was soon shown that through cross referencing with public datasets like phone book listings, individuals and their search preferences could be traced \cite{nyt}. This amounts to breaching the customer's right of privacy. Eventually, lawsuits were filed and AOL let go of the personnel responsible for the crowdsourcing initiative.

\section{Existing approaches to improve Trustworthiness} \label{existingTechniques}
Crowdsourcing can be characterized as a set of individuals who interact with each other as per certain protocols enforced by the platform. The techniques and methods suggested to build trustworthiness of the crowd can be classified into two categories \cite{Ramchurn2004} - a) Individual based approaches and b) System based approaches. 

Individual based approaches focus on choosing the right individual for a task - the assumption being that a trustworthy individual will produce a trustworthy work submission.

System based approaches attempt to build \& enforce a mechanism of interaction between the crowdsourcer and the crowd worker such that the trustworthiness of the work submission can be ascertained. Here, the focus is typically on the work submission and not on the submitter.

We explore each category below, highlighting some of the vendor approaches and the shortcomings of each technique.

\subsection{Individual based approaches}
Individual based approaches attempt to model the characteristics of crowd workers. Characteristics typically include \cite{allahbakhsh2013quality} the reputation, the credentials and the experience. 

The reputation of a crowd worker is what is being said about him from external sources (i.e. not self-proclaimed). Reputation can be computed through different techniques. Evidence based techniques \cite{Huynh2006} compute a score for an individual based on his performance on past tasks. A typical example is the `acceptance rate' \cite{difallah2012mechanical} in AMT where the reputation is computed based on the number of prior task submissions that were accepted. Similar evidence based metrics are followed by Upwork \cite{kokkodis2013have} and TopCoder \cite{topRating} where the reputation for a crowd worker is calculated based on a manual review of the task submission. 

Social relationship based reputation technique has been developed in \cite{sabater2002reputation} where the reputation of the referrer is used to compute the reputation of the referee. Social techniques also include the mechanism to capture open ended comments or testimonies from other stakeholders. These comments help identify a richer set of attributes than the pre-defined dimensions of a reputation score.  

Credentials and experience are typically self-proclaimed (and thus can be lied about). Credentials include being part of a community of professionals (e.g. passed a certification in Java programming) or having executed a legal agreement (e.g. having signed a non-disclosure agreement (NDA); signed an indemnity clause, etc.). TopCoder has pre-defined agreements that the crowd worker has to execute before a task can be given to him. Upwork, in addition to pre-defined agreements also allows the crowdsourcer to add additional agreements and clauses.

Experience refers to the crowd worker's knowledge and skills and is tied to the type of the task. In contrast, reputation is a generic metric \cite{allahbakhsh2013quality} applicable across different types of tasks and skills (like timeliness, communication, etc.).

Most crowdsourcing vendors focus significantly on Individual based approaches. Some of the vendors perform deep background checks for credentials and look to build a `private verified crowd'.

\subsubsection{Problems with the Individual based approaches} 
The key issue with a reputation score is that it represents the characteristics of an individual in a single dimension \cite{Ye}. Tasks in crowdsourcing are of various kinds, needs differing expertise of the crowd and have differing amount of information from the crowdsourcer. In such a scenario a single dimension may not accurately capture the capability of a crowd worker for a new task. For example, a Java programming task which needs the understanding of the Insurance domain may not be well served by a crowd worker who has the best reputation in Java programming.  

Reputation models make the strong assumption that past behavior is a good judge of future performance. However, the performance of a crowd worker may depend on many factors including motivation, social context (what others are doing) and other personal attributes which may never be measurable to build a realistic model. A highly reputed crowd worker can also make genuine errors \cite{allahbakhsh2013quality}. Further, reputation models are susceptible to planned attacks where a malicious crowd worker legitimately builds reputation with the intention to exploit later.

A simulation study was made in \cite{yu2012challenges} where crowd workers with the best reputation were chosen for the tasks. It was found that tasks were being concentrated to a small group of the best crowd workers. This negatively affected the timeliness of submission as the best crowd workers were not available for many of the tasks.

Certain reputation models can be easily manipulated. The example shown in \cite{ipeirotisBlog} depicts how a crowd worker using the AMT platform can easily (and at low cost) achieve the best possible reputation. The technique simply requires the crowd worker to post tasks on AMT and subsequently answer them himself. Similarly, credentials and experience are often unreliable as they are self-proclaimed.

Individual based techniques also assume that the identity of an individual can be accurately authenticated and it is difficult to change identities. 

Despite the ease in subverting Individual based approaches, the techniques are easier to deploy and are applicable across all of the taxonomy elements (see Table \ref{knownTechniques}) making this approach popular among crowdsourcing vendors.

\subsection{System based approaches}
System based approaches focus on building a mechanism through tools \& techniques to improve trustworthiness of the crowd. In contrast to Individual based approaches, the focus is typically on the submission rather than on the submitter.



The risk of poor quality submissions from the crowd (taxonomy element 1 in Figure \ref{tax}) is mitigated in micro-tasks based crowdsourcing engagements through the use of a gold standard. Here, among the tasks that are given to the crowd, a small percentage is solved by the crowdsourcer so that the answers to these tasks are known. The answers are then compared with those that are given by the crowd workers and only those workers who correctly answer these questions are considered to be trustworthy. It has been reported that such gold standards do improve the overall trustworthiness of the submissions \cite{Eickhoff2011}. However, the gold standard tasks need to be designed such that they are indistinguishable from other tasks \cite{difallah2012mechanical}. In the context of crowdsourcing of software development (and other creative tasks), it is nearly impossible to develop a gold standard since the submissions are subjective and cannot be automatically verified - i.e. submissions for a task such as the development of a piece of software cannot be automatically verified to match a gold standard code. Gold standard based methods also truly work when a particular crowd worker attempts multiple tasks (including the gold standard task).

A variant of the usage of a gold standard is the design of tasks in such a way that the characteristics of a valid solution can be verified easily. For example, in the task of development of software, test cases can be designed to check the functional properties of the code. These test cases can be run when submissions are received to verify the adherence to the requirements. In our previous work \cite{dwarakanath2015} we used such a mechanism to quickly discard submissions not meeting the task requirements. However, our experiment showed that writing automated test cases almost equaled the effort of writing the actual code for the task.

Micro-tasks based crowdsourcing have also shown that aggregating answers from multiple submissions can be as trustworthy as the submission from a single expert \cite{Snow2008}\cite{sheng2008get}\cite{eickhoff2013increasing}. Here methods have been devised \cite{hung2013evaluation} such as Majority Voting and Expectation Maximisation where the consensus of a trustworthy submission is derived from multiple submissions. In the context of crowdsourcing software development, it is difficult to aggregate multiple submissions (i.e. how to merge the best aspects of multiple code submissions?). A possible direction to investigate this approach is the use of `recombination technique' \cite{Latoza}. This method explicitly models the crowdsourcing task as a journey with two milestones. At the end of the first milestone, every crowd worker submits his work product individually. At this juncture, the submission of all the crowd workers are shown to each other. The crowd workers are now given the opportunity to improve their solution by looking at others' work. At the end of the second milestone, the improved work product is submitted by the crowd workers. Such a technique was found to improve the overall quality of the submissions \cite{Latoza}.

The work in \cite{dow2012shepherding} evaluated techniques for the review of subjective tasks where gold standards are not applicable. The results showed that self-assessment (where the crowd worker reviews his work based on a structured set of questions) was as efficient as external assessments. 

Each of the techniques (as elaborated above) depend on some form of manual review. TopCoder uses manual reviews where the assessment is done by a select set of crowd workers. Upwork leaves the assessment to the crowdsourcer. 

There are a few tools to address taxonomy elements 1.2 \& 1.3 (`malicious code submission' \& `non-adherence to best practices'). A static code analysis tool such as Checkmarx's CxSAST \cite{checkmarx} can scan uncompiled code and identify security vulnerabilities including malicious code. Certain coding best practices can be detected by tools such as SonarQube \cite{sonarqube}. 

The most prominent method to tackle the timeliness of submission (taxonomy element 2) is to set-up the crowdsourcing task as a contest where multiple crowd workers aim to finish a task within a fixed timeline. TopCoder follows this methodology. Timeliness can also be promoted through monetary benefits \cite{dow2012shepherding}.

Taxonomy element 3.1 (`use of non-compliant licensed software') can be addressed through tools like Black Duck \cite{blackduck}. These tools scan code and identify whether any open source software has been used. TopCoder uses a mechanism called `extraneous check' \cite{topcoderIP} to identify if a piece of software developed for a different crowdsourcer has been re-used. The technique follows the intuition that typically while re-using, the member would not refactor the code or change variable names. The taxonomy element of `Crowd's work being bound by other agreements' can only be addressed through a background check (an Individual based technique) and cannot be solved through system based techniques. Similarly, Taxonomy element 3.2  (`litigation by the crowd') requires careful creation of contracts and other agreements which can only be tackled through Individual based techniques.

%

Crowdsourcing software development typically requires the crowd workers to access enterprise networks (e.g. to check-in the code developed). Network security should be in place to ensure the crowd worker has access only to the required information. Techniques such as a sandbox environment, having an isolated environment for the crowd and a role based access control would help address these aspects (and taxonomy element 4 - `Network Security \& Access Control).

Loss of IP (taxonomy element 5) can be tackled by breaking large tasks into small pieces where each piece gives very little knowledge of the entire application. Ensuring that different crowd workers work on different pieces will help reduce the risk of loss of IP. Other approaches include anonymization and masking. 

Similarly, Loss of data (taxonomy element 6) can be tackled through rigorous masking and breaking the entire dataset into small pieces which individually do not give away sensitive information. 

\subsubsection{Problems with System based approaches}
System based approaches verify the trustworthiness of each submission without having a bias on its origins. While the approaches are difficult to subvert, it is also hard to deploy. There are few other deficiencies.

The biggest trouble with methods that solely depend on the submission is that attributes of the submitter can jeopardize a submission even when the submission is completely robust. The issue of the ownership of a submission when the submitter has made certain obligations unknown to him is an example (Taxonomy element 3). Here, a background check of the submitter is essential to identify such problems. Similarly, the risk of being subjected to litigation cannot be tackled through System based techniques.

System based techniques also require a lot more effort from the crowdsourcer. Setting up review mechanisms was seen to significantly increase the amount of involvement \cite{Stol2014}. Crowd based peer review requires a second round of crowdsourced tasks which increases the overall cost. Setting up contests where multiple crowd workers can submit also exposes the system to risks such as collusion (the case of Henkel in Figure \ref{prilFig}) and social attacks where a crowd worker may demotivate others for personal gain. This aspect of demotivation was observed in our earlier work \cite{dwarakanath2015} where a few crowd workers criticize a crowdsourcing task (typically on the monetary front) through public comments, eventually dissuading others from participating.

We summarize the known techniques against the various risks in Table \ref{knownTechniques}.
\begin{table}[h!]\scriptsize
\renewcommand{\arraystretch}{1.3}
\centering
\caption{Existing risk mitigation techniques.}
\label{knownTechniques}
\begin{tabular}{|c|p{1.3cm}|p{1.4cm}|p{1.4cm}|p{1.9cm}|}
\hline
\multicolumn{3}{|c|}{Taxonomy\cellcolor[gray]{0.9}} 																																														&Individual based approaches\cellcolor[gray]{0.9}					&System based approaches\cellcolor[gray]{0.9}  \\ \hline
\multirow{3}{*}[-5em]{1}& \multirow{3}{1.3cm}[-5em]{Quality of Submission}&Adhering to the requirements										& reputation; credentials; experience	&	peer review; gold standard; recombination; self-assessment	\\ \cline{3-5} 
												&                    															&Malicious code submission  										& reputation													& peer review; static code analysis tools	\\ \cline{3-5} 
												&                   															&Non-adherence to best practices  							& credentials; experience 						& peer review; self-assessment; code analysis tools 	\\ \hline
											2 & \multicolumn{2}{p{2.7cm}|}{Timeliness of Submission} 																									& reputation													& contests \\ \hline
\multirow{2}{*}[-2em]{3}			& \multirow{2}{1.3cm}[-1.7em]{Ownership \& Liability}  		&Use of Non-compliant licensed software	& reputation													& peer review; code analysis tools \\ \cline{3-5} 
												&                   																			&Litigation by crowd										& reputation													& - \\ \hline
											4 & \multicolumn{2}{p{2.7cm}|}{Network security \& Access controls} 																	& reputation													& Intrusion detection systems \\ \hline
											5 & \multicolumn{2}{p{2.7cm}|}{Loss of Intellectual Property} 																							& reputation													& break-down; masking \\ \hline
											6 & \multicolumn{2}{p{2.7cm}|}{Loss of Data} 																															& reputation													& break-down; masking\\ \hline
\end{tabular}
\end{table}

\section{Evidence from Data} \label{evidence} 
In this section we analyzed data from actual crowdsourcing tasks in two prominent vendors - Upwork \& TopCoder. Upwork follows largely Individual based approaches for risk mitigation where a crowd worker is chosen for a task based on his reputation. TopCoder follows predominantly System based approaches where the task to be completed is posted as a contest and any number of crowd workers can participate. The best task submission is chosen as the solution for the task. We look at various metrics to measure the trustworthiness in the two platforms.

The data from the crowdsourcing vendors was collected through public APIs \cite{upworkAPI} \cite{topcoderAPI}. All crowdsourcing tasks that belong to software development were selected. Table \ref{data} shows the details of the data collected. We use `reputation' to denote the score given to a particular crowd worker based on the platform's reputation calculation metric. We use `task score' to denote the review score given to a particular submission from a crowd worker for a task.

\begin{table}\scriptsize
\renewcommand{\arraystretch}{1.3}
\centering
\caption{Data collected}
\label{data}
\begin{tabular}{|p{3.2cm}|p{2cm}|p{2cm}|}
\hline
 \cellcolor[gray]{0.9} & Upwork\cellcolor[gray]{0.9} & TopCoder\cellcolor[gray]{0.9} \\ \hline
 Time period& 1-Mar-2006 to 31-Aug-2015 & 1-Jun-2012 to 31-Dec-2014 \\ \hline
 Num. of tasks completed &86160&7488  \\ \hline
 Num. of task submissions&86160&18195\\ \hline
 Num. of crowd workers&34445&1564\\ \hline
\end{tabular}
\end{table}

\subsection{Amount of Untrustworthy behavior}
Previous research showed that there is a large percentage of untrustworthy workers in the space of micro-tasks. \cite{Eickhoff2011} reported 38\% and \cite{Gadiraju2015} reported 43.2\% of crowd workers were untrustworthy. It has also been thought \cite{Eickhoff2011} \cite{Gadiraju2015} \cite{eickhoff2013increasing} that open ended, complex \& creative tasks are subjected to a lower amount of untrustworthy workers. To verify this claim, we measured the amount of untrustworthy behavior in the macro-tasks of software development. We consider a task submission to be untrustworthy if it's task score is < 75\% of the maximum \cite{archak2010money} (this corresponds to a task score of 3.75 in Upwork and 75 in TopCoder).  
\begin{figure}[!h]
    \centering
    \subfloat{{\includegraphics[width=0.50\linewidth]{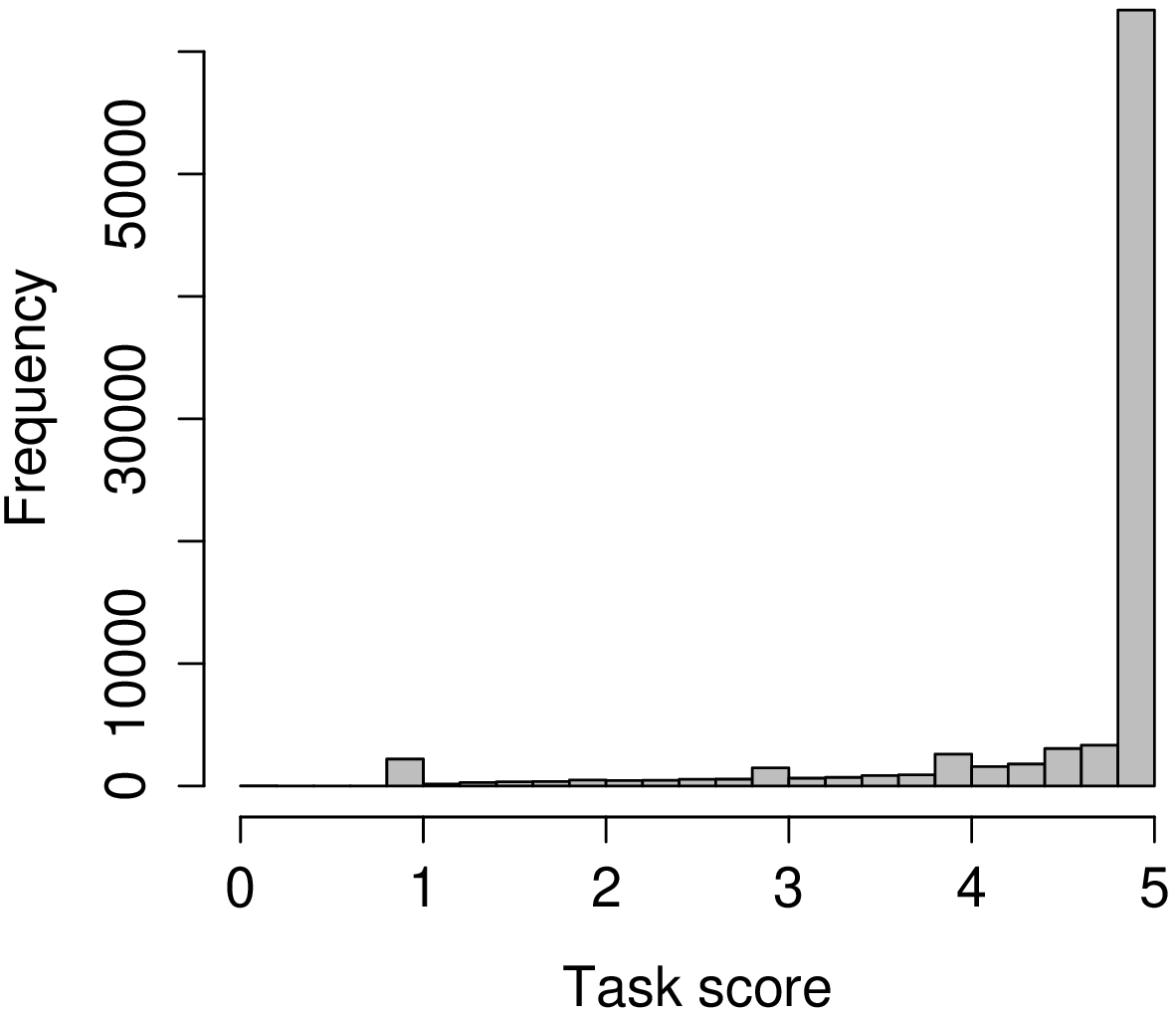} }}
    \subfloat{{\includegraphics[width=0.50\linewidth]{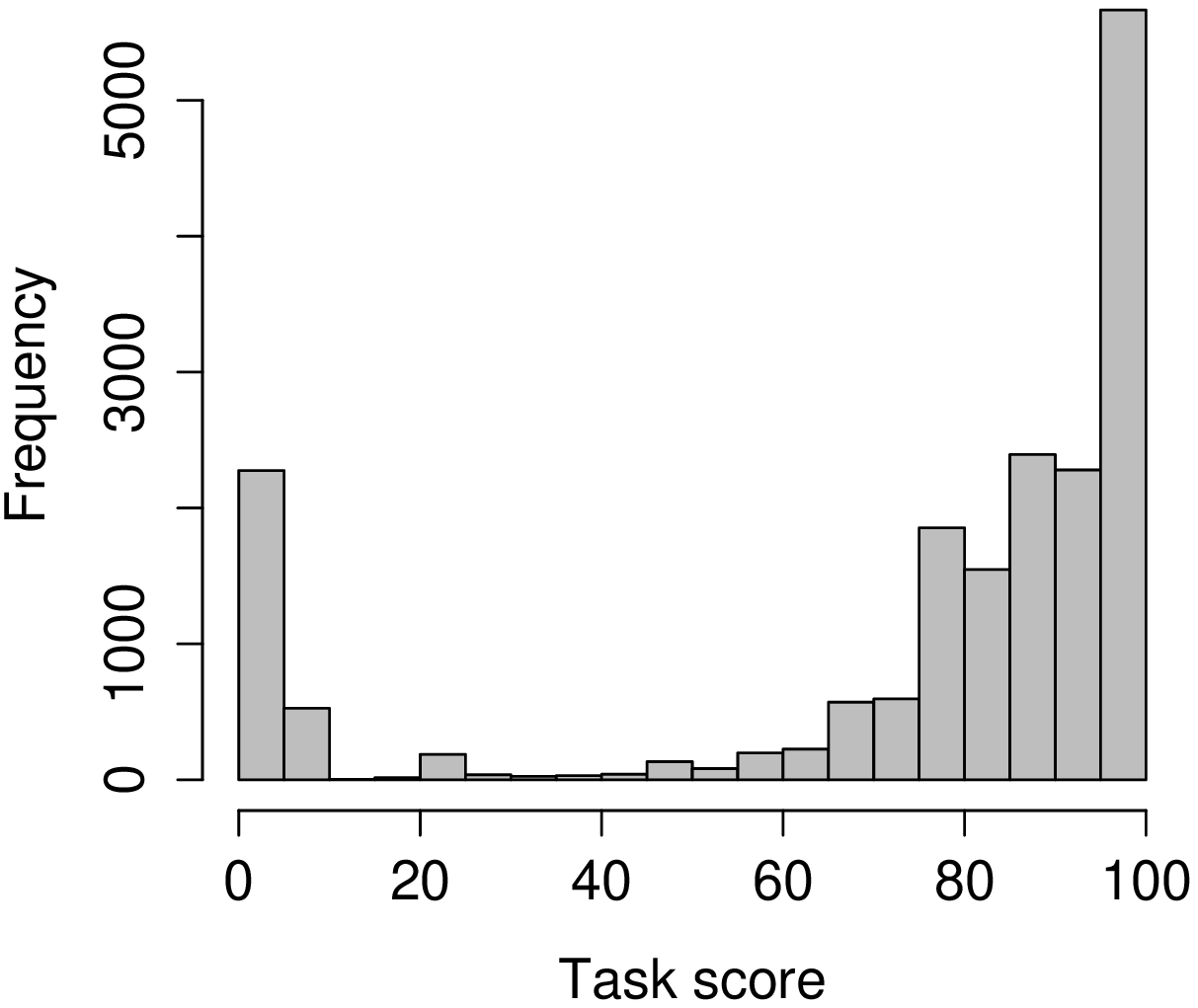} }}
    \caption{Histogram of task scores. a) Upwork b) TopCoder}
    \label{evidence1}
\end{figure}

Figure \ref{evidence1} shows the frequency of task scores. The amount of untrustworthy behavior is 11.5\% in Upwork and 25.89\% in TopCoder. The result for Upwork is significantly less than those reported in the micro-tasks context, while that of TopCoder is marginally less. However, we should note that the task submission in Upwork would have undergone multiple rounds of reviews from the crowdsourcer and corresponding improvements from the crowd worker. The task score reflects the trustworthiness of final submission. In TopCoder the task scores are for the first submission that is made by the crowd workers. We thus believe, the result from TopCoder is a more genuine representation of the amount of untrustworthy behavior in macro-tasks. 

Another interesting observation is that the data in Upwork is extremely skewed: 70.8\% of the tasks received a 100\% task score. This is again a consequence of the hiring mechanism of Upwork where the task score is on the final submission post multiple reviews and edits to code.

\subsection{Reasons for Untrustworthiness}
The data from Upwork provides open ended comments from the crowdsourcers. The comments provide a rich set of data to identify the issues encountered in the crowdsourcing of macro-tasks. There were a total of 1037690 comments in the dataset. We used Latent Dirichlet Allocation (LDA) as a topic model to identify the top set of issues reported by crowdsourcers. This was done to identify if our taxonomy missed any of the top issues. Table \ref{topics} shows the topics from the comments. We found that timeliness of the submission was the most common issue in Upwork. Poor communication skills was also frequently mentioned in the comments. 

Subsequently, we used keywords to search among the comments to look for the occurrence of the risks identified in the taxonomy. The comments and the mapping to the taxonomy is shown in Table \ref{comments}.

\begin{table}[h!]\scriptsize 
\renewcommand{\arraystretch}{1.3}
\centering
\caption{Topics from comments in Upwork}
\label{topics}
\begin{tabular}{|p{0.5cm}|p{3.7cm}|p{2.6cm}|}
\hline
       Task score    \cellcolor[gray]{0.9}        & Top 3 Topics from the topic model \cellcolor[gray]{0.9} & Interpretation \cellcolor[gray]{0.9}\\ \hline
\multirow{3}{0.5 cm}[-3em]{0 to 2} & communication poor quality skills good lack english disappointed articles extremely & Quality of submission / Adhering to requirements \\ \cline{2-3} 
                  & complete completed task job simple tasks requested unable required long & Timeliness of submission \\ \cline{2-3} 
                  & project deadline agreed deadlines due missed meet progress set requirements & Timeliness of submission \\ \hline
\multirow{3}{0.5 cm}[-3em]{2 to 3.75} & job complete task completed time unable needed quickly fine successfully & Timeliness of submission  \\ \cline{2-3} 
                  & deadlines due communication deadline issues meet missed lack availability schedule & Timeliness of submission \\ \cline{2-3} 
                  & good needed skills wasn experience company business set provider level  & Quality of submission / Non-adherence to best practices \\ \hline
\end{tabular}
\end{table}

\begin{table} [!h]\scriptsize
\renewcommand{\arraystretch}{1.3}
\centering
\caption{Actual comments from crowdsourcers.}
\label{comments}
\begin{tabular}{|c|p{1.2cm}|p{1.1cm}|p{3.7cm}|}
\hline
\multicolumn{3}{|c|}{Taxonomy\cellcolor[gray]{0.9}} 																																														&Comment snippets\cellcolor[gray]{0.9}					  \\ \hline
\multirow{3}{*}[-8.5em]{1}& \multirow{3}{1.2cm}[-8.5em]{Quality of Submission}&Adhering to the requirements										& \textit{I have worked with over 50 happy contractors. My requirements are perfect. He is new to oDesk so to get the job he said he will do all required. In the end, he says job is complete and it's 1/4th the task at hand}		\\ \cline{3-4} 
												&                    															&Malicious code submission  										& \textit{I don't know why you tell me that you don't have malicious code. I can see in picture which you sent me, that warning was reported in your Firefox browser. Nice try, I am very disappointed with you and your work}														\\ \cline{3-4} 
												&                   															&Non-adherence to best practices  							& \textit{Overall, not a good working knowledge of MVC design patterns}  							\\ \hline
											2 & \multicolumn{2}{p{2.3cm}|}{Timeliness of Submission} 																									& \textit{Missed the first deadline , promised to make deliver if I gave him extra time , again missed the deadline} 													 \\ \hline
\multirow{2}{*}[-2em]{3}			& \multirow{2}{1.2cm}[-2em]{Ownership \& Liability}  		&Use of Non-compliant licensed software	& \textit{His works meets my requests. However he removed copyrights from third-party Apache2.0 Licensed source code. It is license violation}												 \\ \cline{3-4} 
												&                   																			&Litigation by crowd										& - \\ \hline

											4 & \multicolumn{2}{p{2.3cm}|}{Network security \& Access controls} 																	& \textit{Don't trust this contractor.  Attempted to change site owner access and hack server}												 \\ \hline
											5 & \multicolumn{2}{p{2.3cm}|}{Loss of Intellectual Property} 																							& \textit{He signed Non Disclosure Agreement with us, but he dare to display the mockup homepage of our project website in his portfolio. This is illegal and violate our intellectual property}													 \\ \hline
											6 & \multicolumn{2}{p{2.3cm}|}{Loss of Data} 																															& \textit{All we asked for, is, please do not use our asset that were send to you for your own or any other purpose, either commercially or non commercially}													 \\ \hline
\end{tabular}
\end{table}

\subsection{Effect of Geography on Trustworthiness}
\begin{figure} [h!]
    \centering
    \subfloat{{\includegraphics[width=0.50\linewidth]{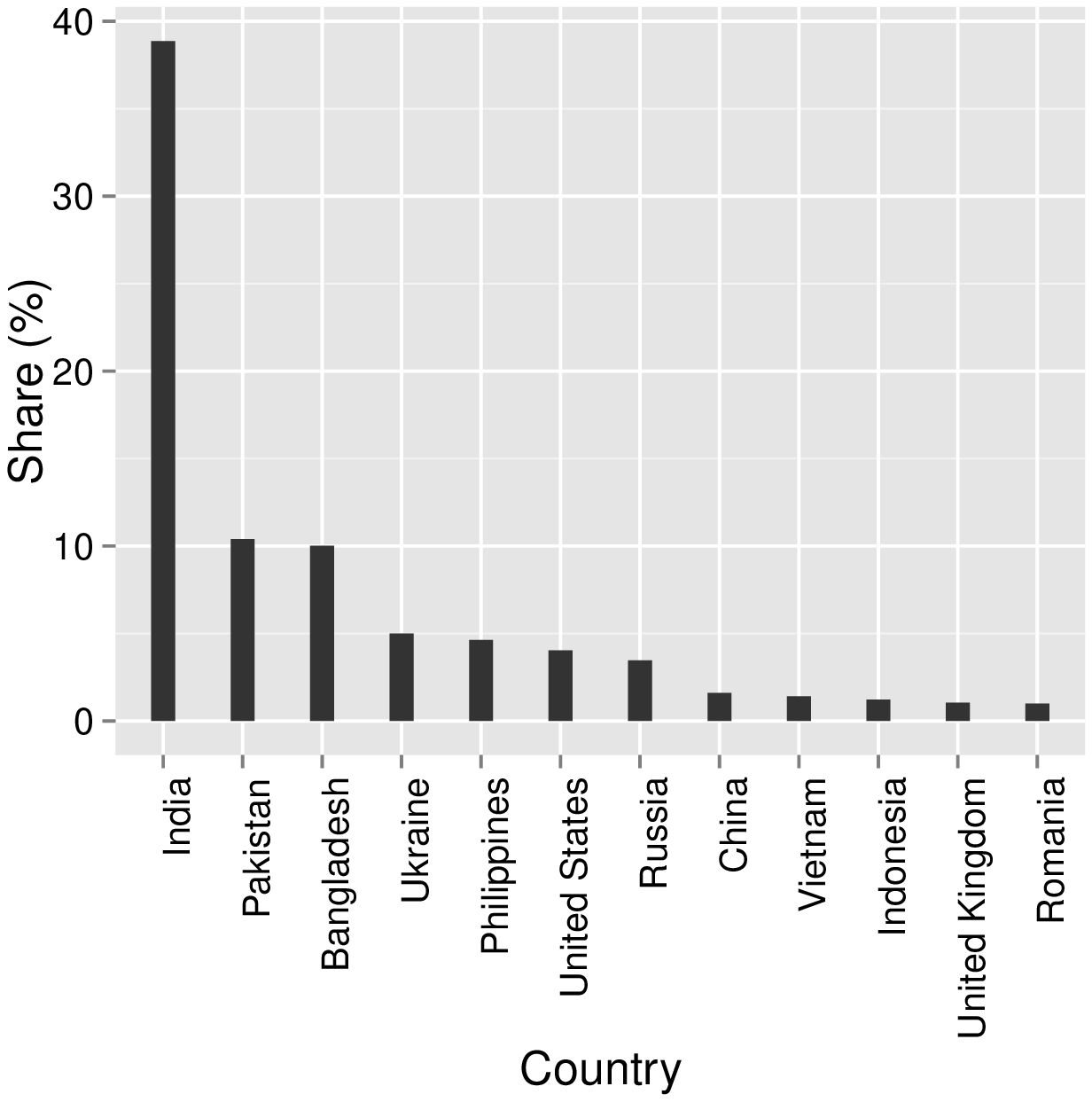} }}
    \subfloat{{\includegraphics[width=0.50\linewidth]{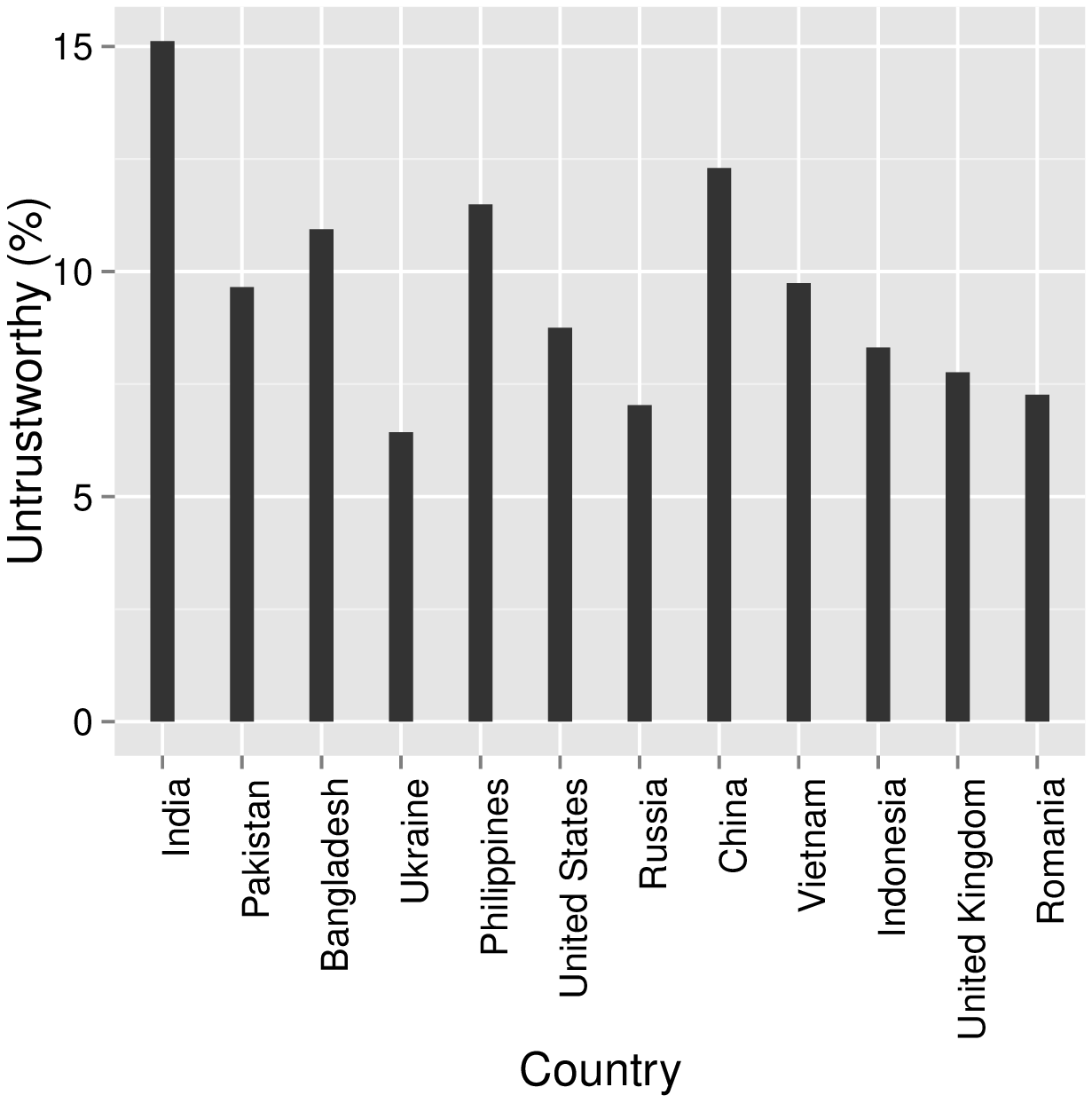} }}%
    \caption{Geographical distribution in Upwork. a) Crowd workers per region. b) Untrustworthy \% per region}
    \label{evidence4u}
\end{figure}

\begin{figure} [h!]
    \centering
    \subfloat{{\includegraphics[width=0.50\linewidth]{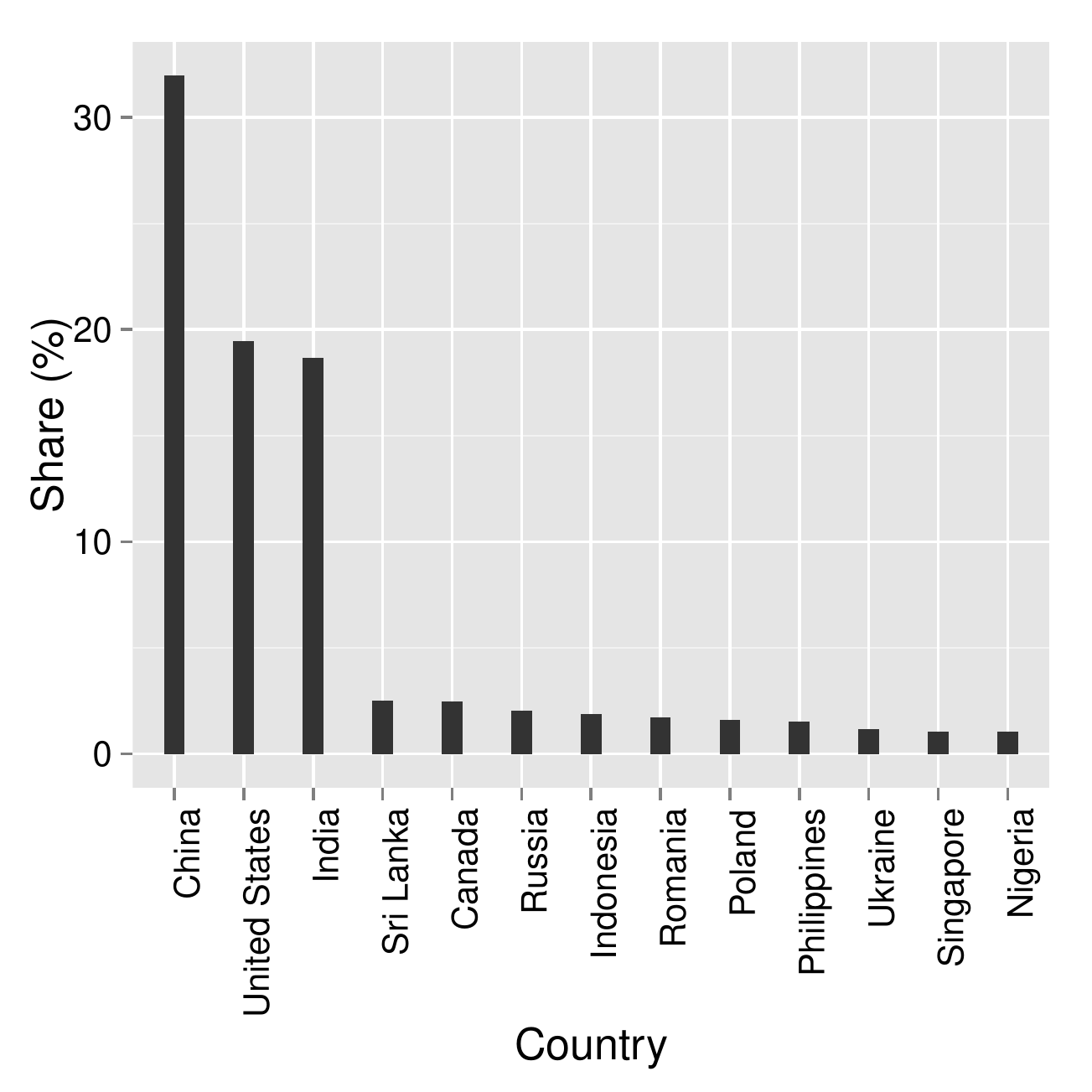} }}
    \subfloat{{\includegraphics[width=0.50\linewidth]{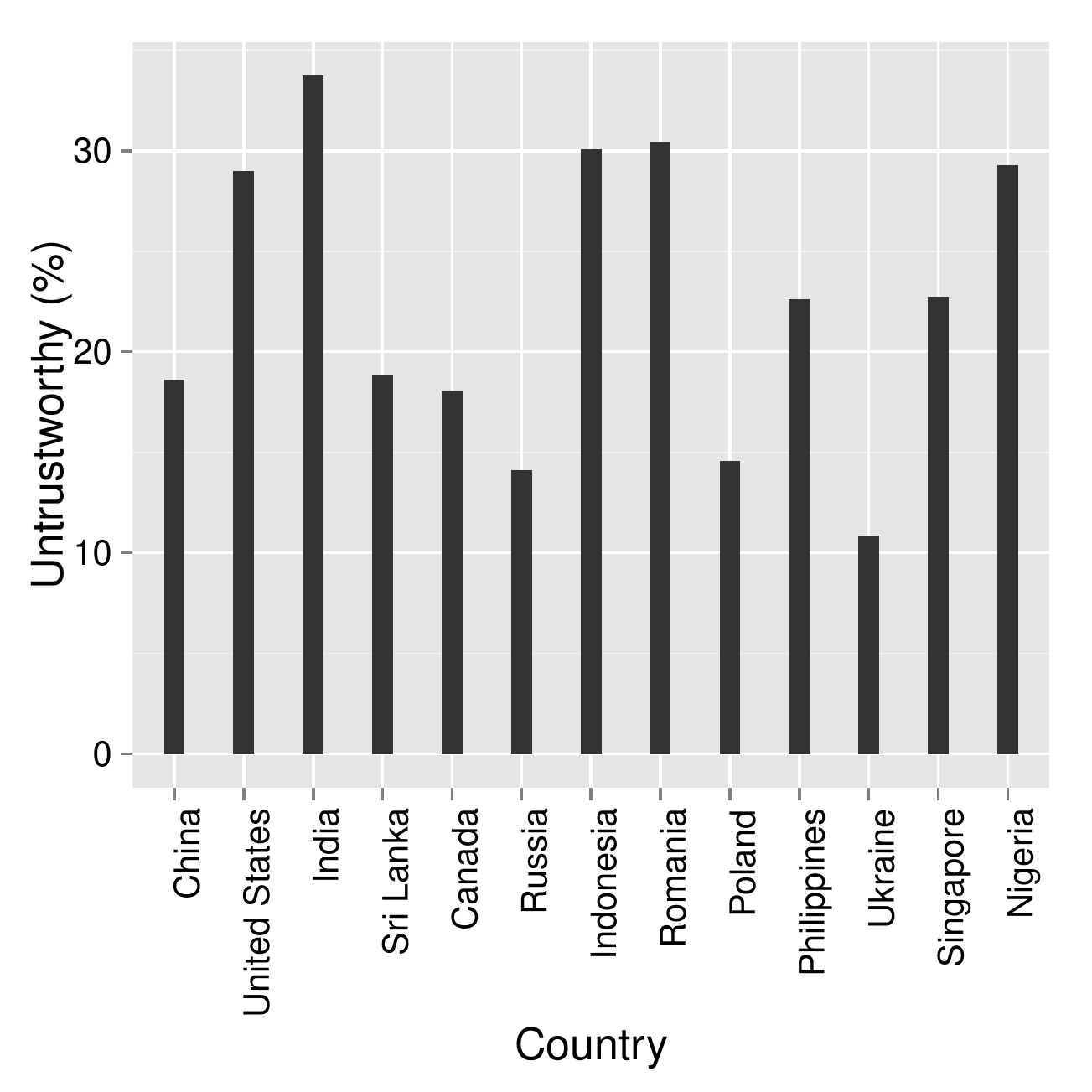} }}
    \caption{Geographical distribution in TopCoder. a) Crowd workers per region. b) Untrustworthy \% per region}
    \label{evidence4t}
\end{figure}

The overall share of crowd workers across the top geographies and the amount of untrustworthy behavior across these geographies are shown in Figures \ref{evidence4u} and \ref{evidence4t}. Most of the crowd workers were from India (37\%) in Upwork and China (32\%) in TopCoder. The amount of untrustworthy behavior varied across the regions with India topping the amount of untrustworthy crowd workers in both Upwork and TopCoder.

The work in \cite{Eickhoff2011} and \cite{Gadiraju2015} reported that choosing crowd workers based on geography makes a significant difference on trustworthiness for micro-tasks. To check whether this holds true for macro-tasks, we performed a statistical hypothesis test.  We ran the Welch two sample single sided t-test with null hypothesis that the population of crowd workers from the `U.S.' geography has a higher average task score than the entire population. We also employed the Cohen's d metric to measure the effect size. Effect size gives a measure of the significance of the difference between the two populations. 

For Upwork the p-value was 1.908e-12 and the effect size was 0.10. This result shows that choosing the crowd workers based on the `U.S.' geography does increase the mean task score (i.e. null hypothesis is rejected), however the effect is small to be practically beneficial (i.e. small effect size). 

For TopCoder, the p-value was 1 (i.e. null hypothesis cannot be rejected) which indicates that choosing crowd workers from the `U.S.' geography does not improve trustworthiness. Thus, we observed that geography does not make an impact on the trustworthiness for macro-tasks.

\subsection{Effect of monetary benefits on Trustworthiness}
\begin{figure} [h!]
    \centering
    \subfloat[All tasks. $r = 0.0060$]{{\includegraphics[width=0.5\linewidth]{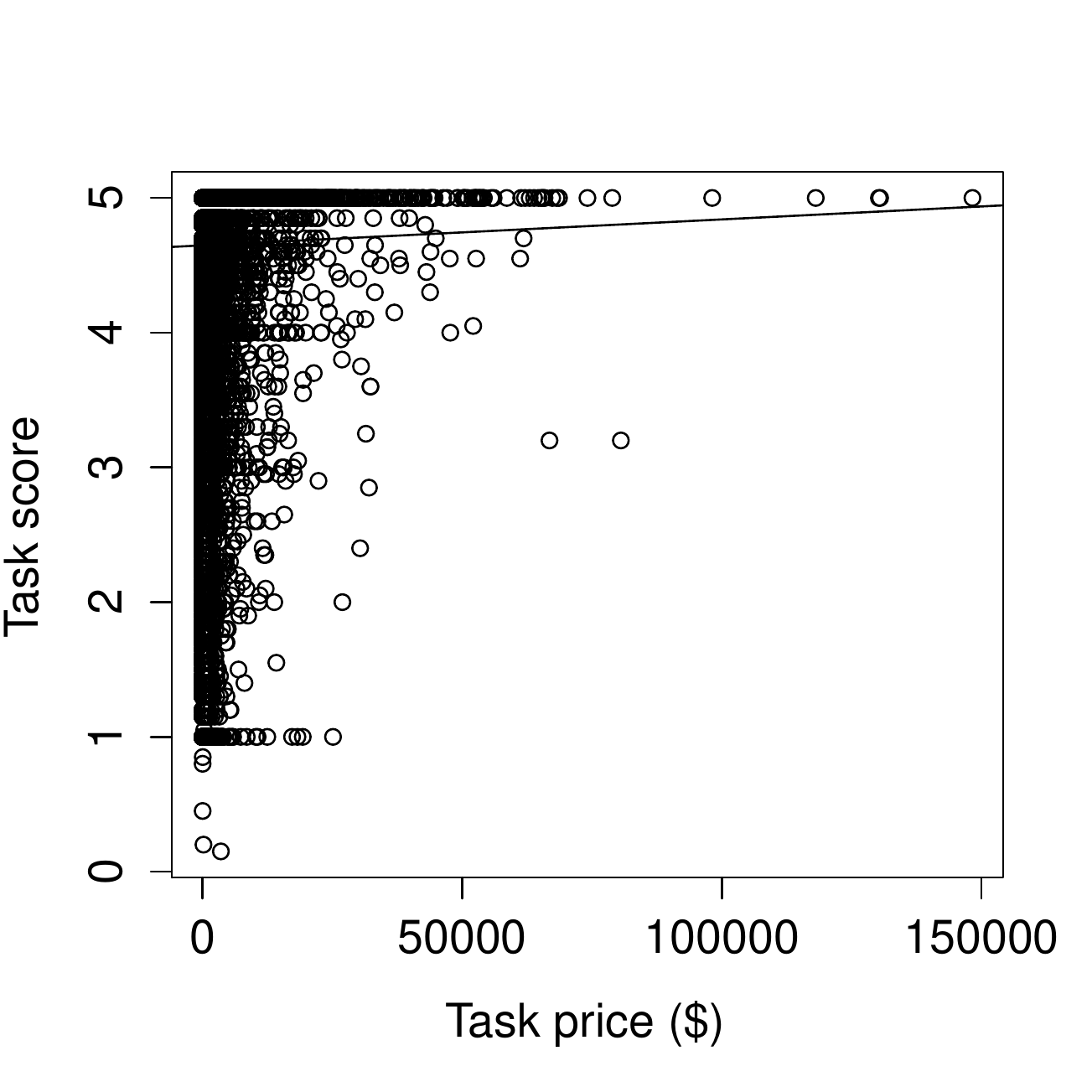} }}
    \subfloat[Tasks with prize <= \$97.23. $r= 0.0213$]{{\includegraphics[width=0.5\linewidth]{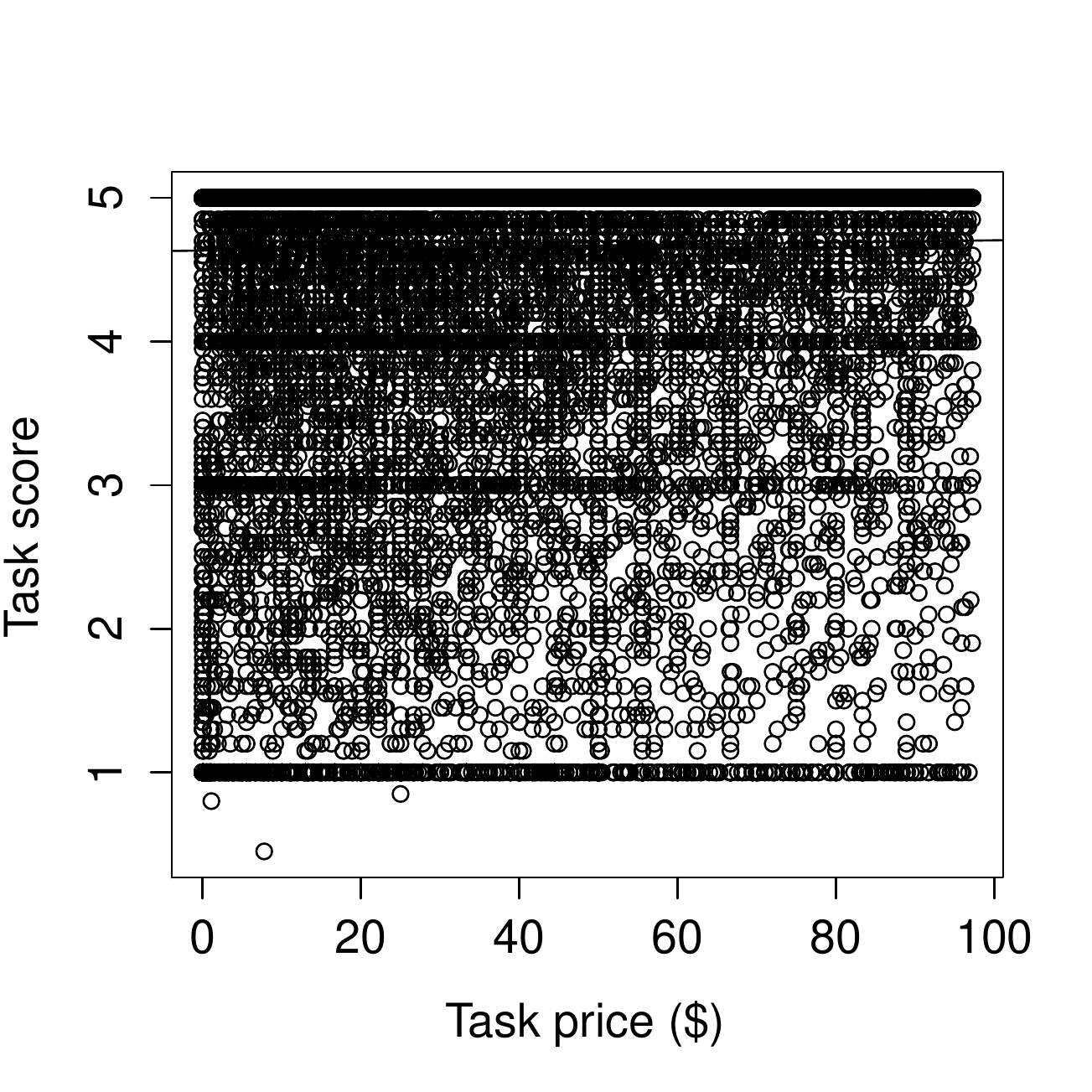} }}
    \caption{Task score versus monetary benefits in Upwork.}
    \label{eveidence5u}
\end{figure}

\begin{figure} [h!]
    \centering
    \subfloat[All tasks. $r= 0.0884$]{{\includegraphics[width=0.5\linewidth]{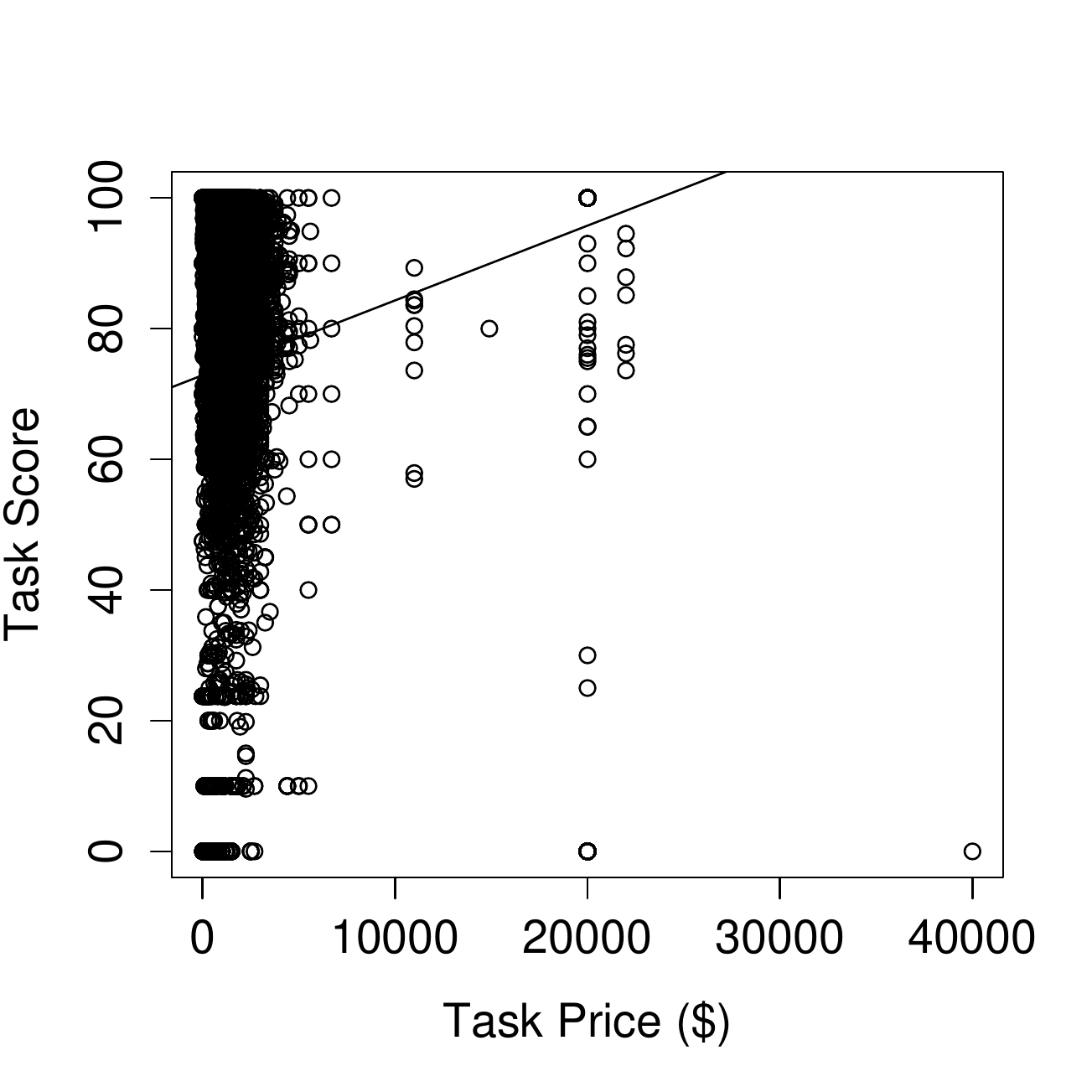} }}
    \subfloat[Tasks with prize <= \$900. $r= 0.2551$]{{\includegraphics[width=0.5\linewidth]{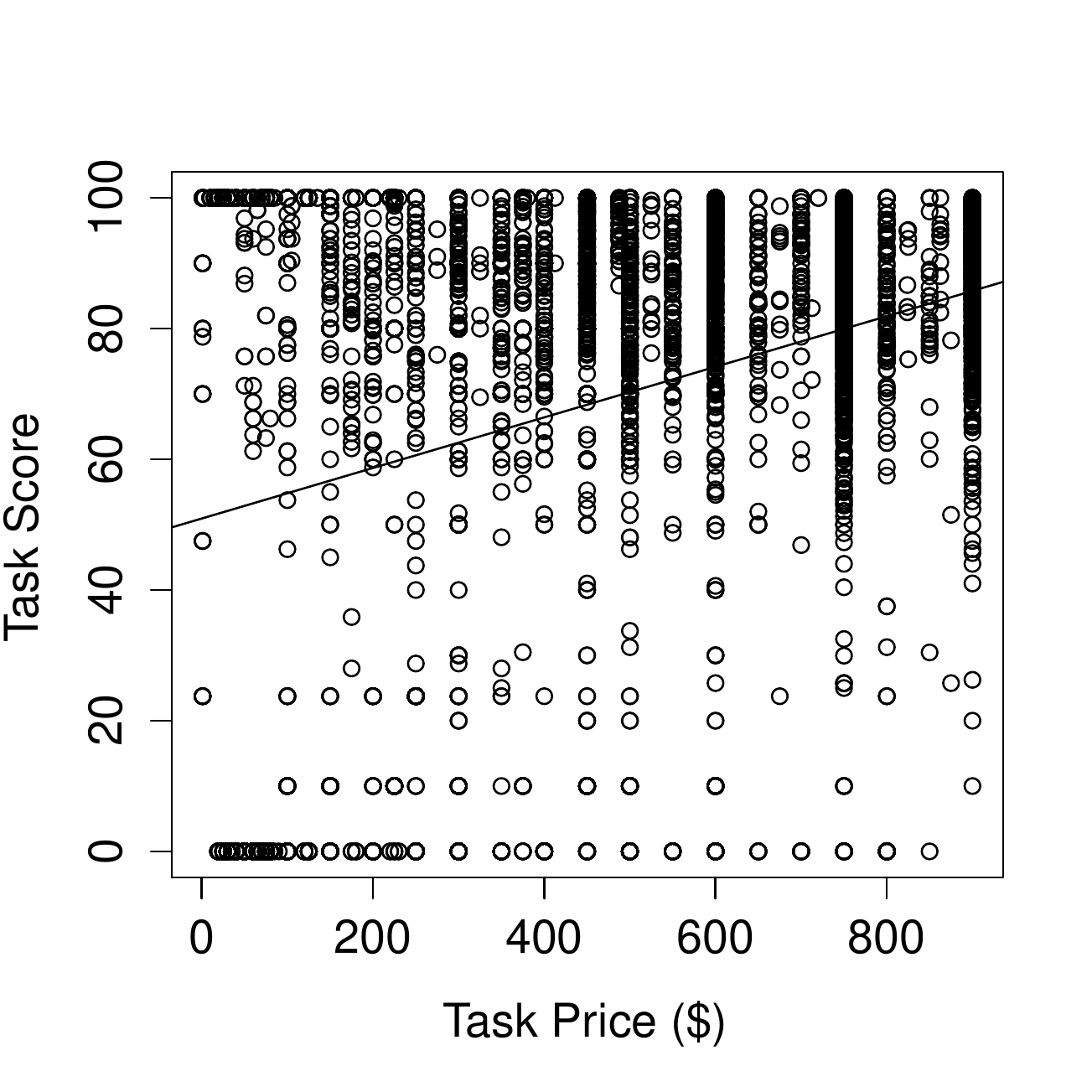} }}
    \caption{Task score versus monetary benefits in TopCoder.}
    \label{eveidence5t}
\end{figure}

We performed a linear regression between the task score and the monetary benefits of a task. The correlation co-efficient was close to 0 (refer Figures \ref{eveidence5u} and \ref{eveidence5t}). This finding is similar to the effect of monetary benefits as reported in \cite{dow2012shepherding} and \cite{mason2010financial} for micro-tasks. As the range of the prize was large, we also performed the regression till the median prize. The median for Upwork was \$97.23 \& for TopCoder was \$900. 

\subsection{Effect of Reputation on Trustworthiness}
\begin{figure} [h!]
    \centering
    \subfloat{{\includegraphics[width=0.50\linewidth]{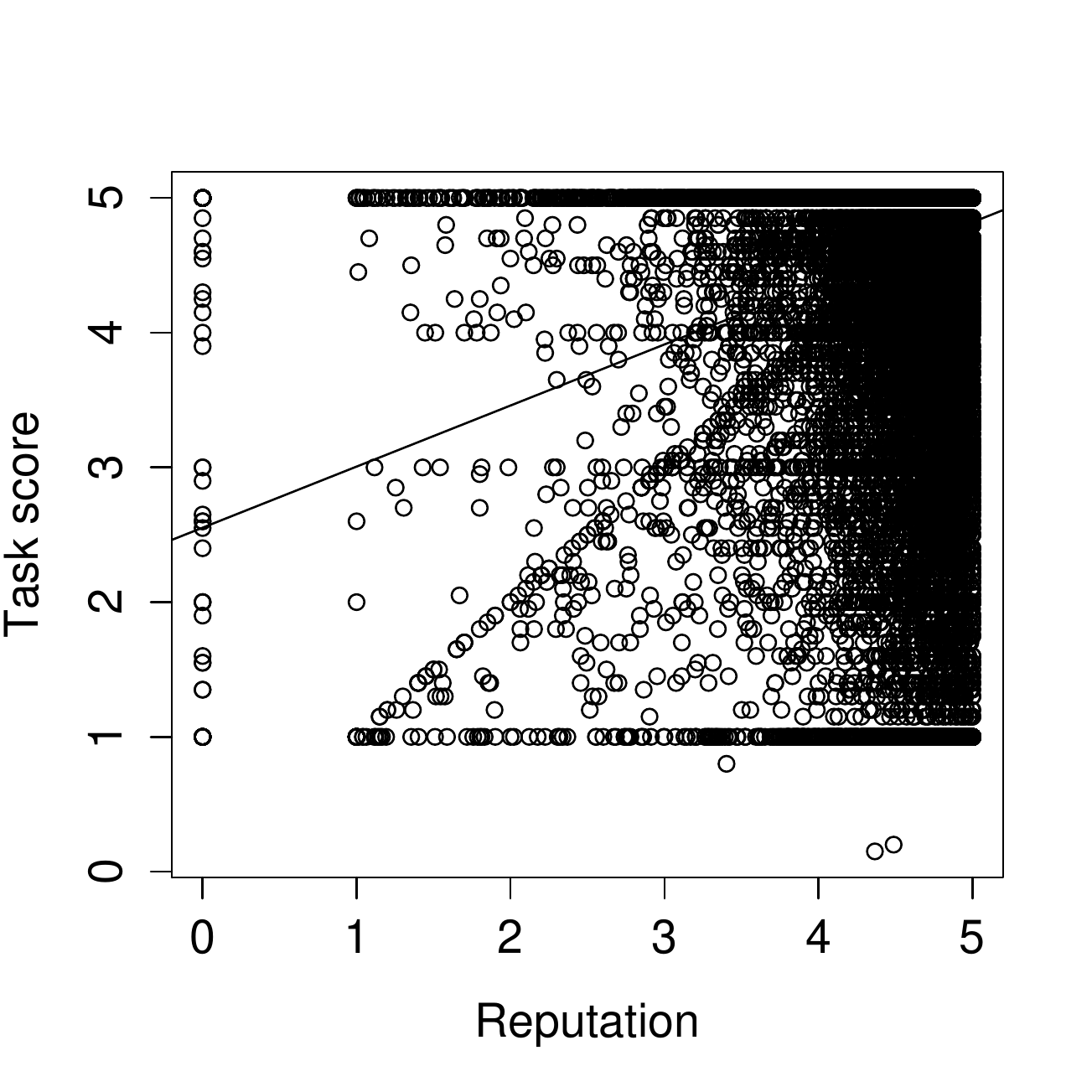} }}
    \subfloat{{\includegraphics[width=0.50\linewidth]{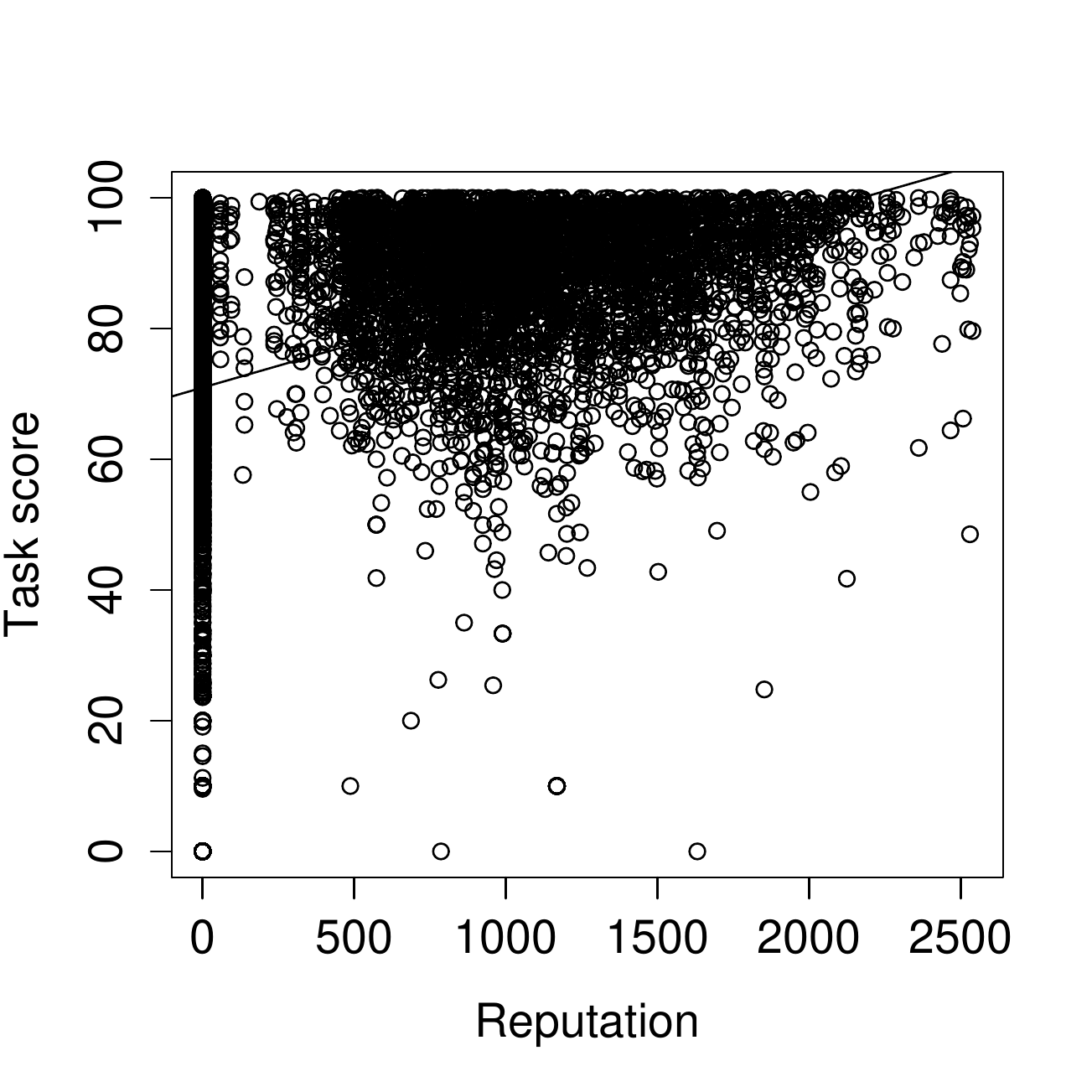} }}
    \caption{Task score vs. reputation. a) Upwork b) Topcoder}
    \label{evidence3}
\end{figure}

Reputation is a type of Individual based approach and makes the assumption that a crowd worker with a better reputation is more trustworthy. Upwork predominantly follows this approach to improve trustworthiness. We tested the performance of the method in two ways. First we performed a linear regression between the task score and the reputation. The correlation co-efficient indicates the influence of reputation on the task score. Figure \ref{evidence3} shows the plot of task score versus reputation. The correlation co-efficient was 0.2457 for Upwork and 0.2072 for TopCoder, indicating that task score and reputation are almost uncorrelated.  

Because of the skewness of data (where almost all crowd workers have a high reputation in Upwork - see the dense plot in Figure \ref{evidence3} a) ), we performed a statistical hypothesis test to check whether choosing crowd workers with a high reputation produces a better task score than choosing a crowd worker at random. We ran the Welch two sample single sided t-test with null hypothesis that the population of crowd workers with a high reputation has the same mean of task score as the entire population. Effect size was measured using Cohen's d.

For Upwork, we tested for crowd workers with a reputation of >=4.5. The t-test gave a p-value of 2.2e-16 and the effect size was 0.076. This indicates that choosing crowd workers with a high reputation does increases the task score (i.e. the null hypothesis can be rejected) however practically it would make marginal difference (i.e. small effect size). 

We also checked the amount of untrustworthy crowd workers in Upwork who have a reputation of >=4.5. This came to 6\%. Thus, by choosing a highly reputed crowd, while the overall quality of submissions does not significantly increase, the chance of finding poor results reduces (i.e. the mean is almost the same but the variance decreases). 


\subsection{Effect of Contests on Trustworthiness}
TopCoder posts crowdsourcing tasks as contests without typically restricting who can participate. We tested the performance of this System based approach on trustworthiness. 

We selected all contests that were posted and chose the submission that had the maximum task score for that contest. This task submission is considered as the solution for the task. We then verified the amount of untrustworthy solutions for the contests - i.e. the number of contests where the best submission has a task score of < 75. This resulted in the amount of untrustworthiness of 3\%. The reduction is significant where the usage of contests has reduced the amount of untrustworthy behavior from 25.8\% to 3\% and strongly indicates the usefulness of contests.

%


\subsection{Summary of the empirical tests}
The empirical tests have provided a few key results. 
\begin{itemize}
\item Unlike previously thought (\cite{Eickhoff2011}\cite{Gadiraju2015}\cite{eickhoff2013increasing}), we found a sizable amount of untrustworthiness in macro-tasks.
\item We found strong statistical evidence showing that geography does not make an impact on trustworthiness for macro-tasks (unlike reports \cite{Eickhoff2011} \cite{Gadiraju2015} for micro-tasks).
\item Similar to earlier research (\cite{dow2012shepherding} \cite{mason2010financial}), we found poor correlation between trustworthiness \& monetary benefits.
\item For the first time, we found strong statistical evidence for the poor impact of reputation on trustworthiness.
\item For the first time, we found evidence of the strong efficacy of contests in reducing untrustworthy behavior.
\end{itemize}

While our results are based on statistical evidence, there are a few threats to validity. The biggest aspect is the subjective nature of task reviews. In Upwork, task scores are given by different crowdsourcers and thus two task scores may not be comparable (i.e. crowdsourcer A's review of 4/5 may be equivalent to crowdsourcer B's score of 2/5). Further, for this precise reason, task scores across Upwork and TopCoder cannot be directly compared as well.

The skewness of data may limit the accuracy of t-test which assumes a normal distribution for the data samples.

While the performance of contests looks appealing, our tests have not been able to capture the costs (prize money) involved. As seen in Figures \ref{eveidence5u} and \ref{eveidence5t}, the range of monetary benefits is significantly different between Upwork and TopCoder. This difference could play an important role in trustworthiness to compare the platforms.

Similarly, we have not been able to measure the aspect of the timeliness of submission in contests versus reputation (data was not available). In our previous work \cite{dwarakanath2015} we noticed that contests have a poorer record in timeliness of submission, where no crowd worker attempts a contest.

\section{Conclusion} \label{conclusion}
Our work in this paper studied the aspect of trustworthiness in crowdsourcing in the context of software development. We presented the taxonomy of  trustworthiness and the existing methods to build the trust in the crowd. We also studied empirically the impact of certain mitigation techniques on trustworthiness. Our study and results can serve as a guideline for enterprise crowdsourcing. 

Individual based approaches (particularly for `Ownership \& Liability') and System based approaches (for run-time validation) are essential for a robust crowdsourcing initiative. The impact of reputation was found to be low and thus focusing on building a `private verified crowd' is possibly only part of the solution. Our results also indicated the efficacy of contests as a mechanism to improve trustworthiness.

We believe our comprehensive taxonomy, survey of existing techniques and the results from the large empirical analysis will give impetus for the adoption of crowdsourcing in an enterprise context.


%

\bibliographystyle{abbrv}
\bibliography{references}  
%
%
\end{document}